# Structure prediction of two-dimensional materials based on neural network-driven evolutionary technique.

K. Zberecki,

Warsaw University of Technology, Faculty of Physics, 00-662 Warsaw, Poland

krzysztof.zberecki@pw.edu.pl

**Abstract**

We present a simple yet effective method for structure prediction of two-dimensional structures. The method is based on a combination of neural network and evolutionary techniques. It allows finding pristine 2D structures as well as structures grown on a substrate. Conducted tests show, that the method is efficient and the calculations, based only on the information of stoichiometry, can lead to stable structures. Since the algorithm is able to address structures on a given substrate, it can be useful from the experimental point of view.

**Introduction**

These days, making an attempt to find a perfect combination of chemicals suitable for a specific applications, a chemist (or a material engineer) has two options. The first one is to conduct an experiment, synthesize a sample, and then characterize it with proper methods. The second option is to use a theoretical study first, which may reveal properties of a given material with very high credibility. This is possible mainly due to the huge success of the *ab-initio* methods, based on Density Functional Theory (DFT) [1], which is currently one of the main computational tools for materials sciences, used at the atomic level. The increasing computational power and data storage capacity have made possible to calculate and analyze properties of a huge amount of materials. That high throughput (HT) studies (e.g. [2]) are becoming more and more popular because at relatively low cost allow to pick out structures with desirable properties from a large group of average ones. Moreover, a large amounts of generated data ignited studies based on methods used earlier in pure data science, such as deep learning methods [3], usually based on artificial neural networks (NN) [4] and evolutionary optimization methods [5]. An additional effect of these HT studies is the creation of free-accessible on-line databases [6-8], containing thousands of calculated structures.

From the implementation point of view of the previously mentioned methods, one may observe three main directions. First one is the application of the methods based on evolutionary algorithms, as implemented for example in `XtalOpt` [9] or `USPEX` codes [10]. Especially the latter was extremely successful, being used in more than half a thousand studies. The second direction one may take is a swarm optimization [11], implemented for example in the `CALYPSO` code [12]. A third way is the use of (more and more popular in all fields of science) artificial neural networks. These studies take advantage of the large amount of available data to train the network and then use it to predict properties of yet unknown structures, for example for semiconductors [13], perovskite crystals [14], molecular crystals [15] or molecules [16]. Moreover, according to [17], it is possible to build a NN model, which predicts crystals' properties with greater accuracy (with respect to an experiment) than a Hybrid DFT method does. One of the disadvantages of this approach, though, is the necessity of having the proper set of data, on which the NN would learn. Also, of course, this is in fact only a phenomenological approach.

It also possible to combine methods belonging to one of the mentioned paths. Especially promising (e.g. in terms of performance) seem to be methods combining NN and evolutionary approach [18]. In these hybrid methods, NN and DFT evaluations of the total energy are used in the pursuit of the optimal structure.

Low-dimensional materials play an increasingly important role in the pursuit of the new generation of structures that will build future logic systems. Extensive theoretical work allowed to get a huge number of potentially valuable substances. Similarly, significant progress has been made in the experimental field [19]. Also, very recently, the first two-dimensional (2D) magnetic material has been experimentally confirmed [20]. These materials are not only flat as graphene but may also be build of several monoatomic layers [19]. Also, experimentally, a 2D layer is always placed on a substrate, so theoretical models have to include its impact.

In this paper, we present a simple, yet effective algorithm of finding new two dimensional structures, either in vacuum or on a substrate, starting only with its stoichiometry. Since the main issue of the new materials search problem is a vast search space [10], to effectively reduce this space, the algorithm combines trained NN and an evolutionary approach.

**Description of the used algorithm**
This section presents the detailed algorithm of an approach, that we developed to predict properties of 2D structures.

I. Applied methods
Generally, the algorithm consists of three components.
The first one is an artificial neural network model, prepared (i.e. taught) with a collection of existing data. Here, the data from `c2db` [8] database has been used. This database contains calculated data of more than 3800 different 2D structures, including atomic, electronic, and magnetic properties. It is then a good set for training of a NN. This model, when given a new, unknown structure (i.e. not present in the database) as an input, is able to give as an output two estimations: the total energy of a unit cell and/or its lattice constants (accuracy of these estimations is addressed in the 'Tests' section).
The second component is the genetic algorithm. In a typical algorithm of this type, the initial population is generated, then the fitness function of each individual is evaluated and the new population is generated, based on evaluation results. To generate a new population, a number of genetic operations is applied, mostly crossing and mutation [21]. Here, similar to other evolutionary codes [9,10], the operations are carried out on 'individuals', being atomic structures.
As an evaluation tool, and the third component, the `VASP` *ab-initio* code [22] with PAW pseudopotentials [23] has been used.
The whole algorithm has a modular composition – one can test different NN model, based on different data set, as well as to adapt a different set of genetic operations. Also, different *ab-initio* code may be used as a total energy calculator.

II. The algorithm
The algorithm starts with the generation of the initial population. The only input is the stoichiometry of the structure and the population size (`Npop`). Based on the atomic composition, the initial lattice constants are estimated by the NN model. Then, each structure (i.e. unit cell) is created in one of two ways:
a. only randomly
b. randomly, but having (also randomly chosen) specific symmetry group
Each structure is drawn given number of times (default value is 100), then its total energy is estimated by the NN model. The best from each draw is taken to the population. This generates the population initially evaluated, and the numerical cost is a fraction of what would be needed using an *ab-initio* code.

The next step is the *ab-initio* evaluation, taking the total energy of a unit cell as the only fitness parameter. Based on these fitness values, the next population is generated, with the following genetic operations.

a. (Npop/2)-1 least adapted (i.e having the highest total energy) structures are removed from the population

b. The best individual is kept intact

c. Npop/4 new structures are generated by the atoms switch operation

d. Npop/4 new structures are generated by the softmutation operation

e. (Npop/2)-1 new structures are generated randomly, with the same procedure as for initial population

Atoms switch is an operation, in which a 'good' structure (i.e. better than at least the other half ones in the current population) is chosen randomly and two, also randomly chosen atoms are switched, generating new structure with the same lattice vectors. Softmutaion [10] is an operation in which the coordinates of one of the atoms of a 'good' structure are modified so the structure's total energy is lowered. For this operation, the NN model is used again for energy estimation, and the simple gradient descent method is used for minimization. As one can see, there is no heredity operation used. As our tests with the USPEX code revealed, in the case of 2D materials, the individual which minimizes the total population is rarely an effect of heredity, and rather of softmutation.

The algorithm, shown in Fig. 1, stops when the given total number of generations is reached or the best individual prevails given number of generations.

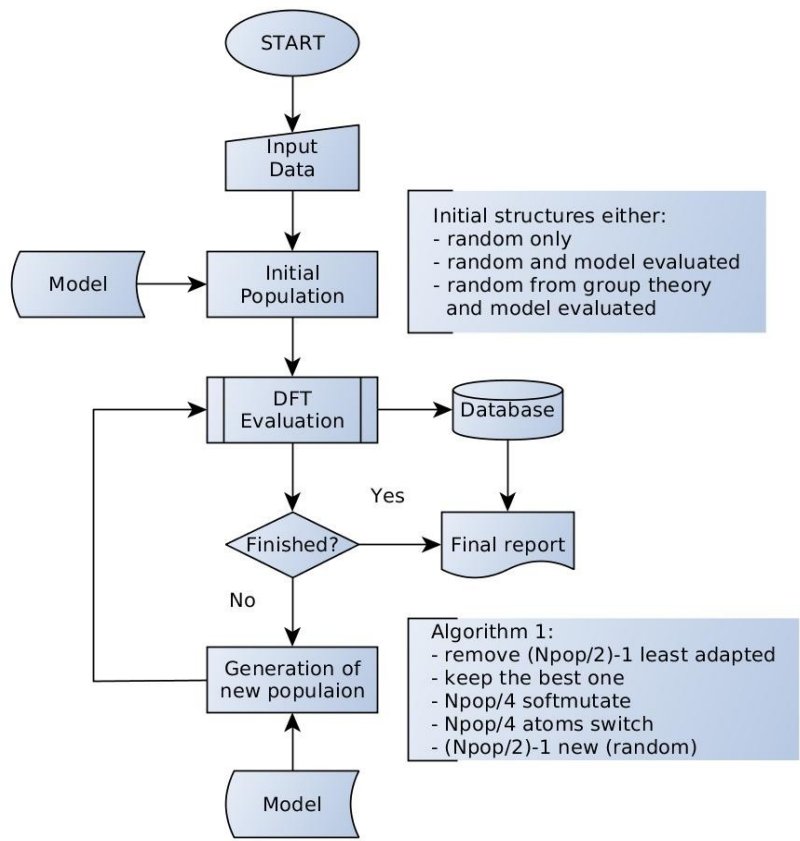

Fig. 1 Flowchart of the algorithm of described method.

**Implementation**

The above algorithm has been implemented in the `NGOpt` code (Neural network Genetic algorithm Optimizer), written entirely in the Python programming language [46]. The code uses a few external free libraries.

The main data structure is the `Individual` class, which inherits from the `Atoms` class of the `Atomic Simulation Environment` (`ASE`) [24] and describes a distinct structure. The initial population (a set of `Individuals`) is prepared randomly, with the use of the `PyXtal` library, which allows one to generate random crystal structures with given symmetry constraints [25]. The NN model used for the pre-evaluation of structures (and also for softmutation) has been prepared with the use of the `MEGNet` library [27], which implements the DeepMind's graph networks [26]. Thanks to its internal structure, being a sequence of traditional NN layers and the graph networks, `MEGNet`-based models can achieve very low prediction errors in a broad array of structural properties. `NGOpt` uses `MEGNet` model (trained on `c2db` data) to predict the total energy of a given structure. Predictions of the lattice constants are made with a simpler model, based on `Keras'` library [28] dense neural network class (trained on the same dataset). The code uses `VASP` for *ab-initio* structure evaluation, although has been successfully tested with the `Siesta` [29] and `OpenMX` [30] programs.

**Tests and examples**

As examples three systems have been chosen, each having a little different nature. First, though, the accuracy of the NN model should be examined.

I. The NN model

While testing the artificial neural network model, one usually divides the dataset in two, the first set is used for training while the other for testing. Here, we used 3200 2D structures from the `c2db` for training and 600 structures for testing. Since the first `NGOpt` test was planned for $MoS_2$, all the data for transition metal dichalcogenides (TMDs) in 1T and 2H conformations were contained in the testing set. The `MEGNet` network with the default number of layers (three `MEGNet` blocks) has been trained for 1000 epochs taking the total energy per atom ($E_{tot}$/at) as a value to be estimated. The results for a few chosen dichalcogenides can be read from the Tab. 1., compared to the database values and our VASP calculations [31].

|         | NN model [eV/atom] | c2db [eV/atom] | VASP [eV/atom] |
|---------|--------------------|----------------|----------------|
| $MoS_2$  | -7.32              | -7.37          | -7.29          |
| $MoSe_2$ | -6.72              | -6.76          | -6.69          |
| $MoTe_2$ | -6.21              | -6.17          | -6.16          |
| $WS_2$   | -7.38              | -7.42          | -7.46          |
| $WSe_2$  | -6.69              | -6.75          | -6.79          |

Table 1 Comparison of the total energy per atom between NN model, `c2db` database data and `VASP` calculations for five most known TMDs.

Comparing all the estimated values of $E_{tot}$/at from the dataset with the actual database values gave an accuracy of 98%, so not the "hybrid DFT quality", but enough for the model to be a reliable estimator in the algorithm. This is mainly due to the fact, that only more than a 3200 structures have been used in the training set, while in the original `MEGNet` project this pool was nearly 70 thousand [26].

II. MoS$_2$ structure prediction

As a first test, the stoichiometry of the most known TMD has been chosen, which is MoS$_2$. We have used VASP as a calculator, the number of generations has been set to 10, and the number of individuals in each population equal to 24 and 48. The results can be seen on Fig. 2, where the energy of each individual in each generation is shown against the total energy per atom. In the case of 24 individuals per generation, the MoS$_2$-2H structure (symmetry group D$_h$) appeared in the structure pool in the 3$^{rd}$ generation (arrow on Fig. 2a), and prevailed until the calculation ended. In the case where the number of individuals in one generation is doubled (Fig. 2b), the 2H structure has been obtained even earlier (generation no. 2). Tests conducted for other TMD from Tab. 1 gave similar results. These simple tests show that the method works and that for simple stiochiometries the convergence may be obtained very fast. Moreover, each calculation can be done on a standard desktop 8-core PC in a 1.5 hours tops.

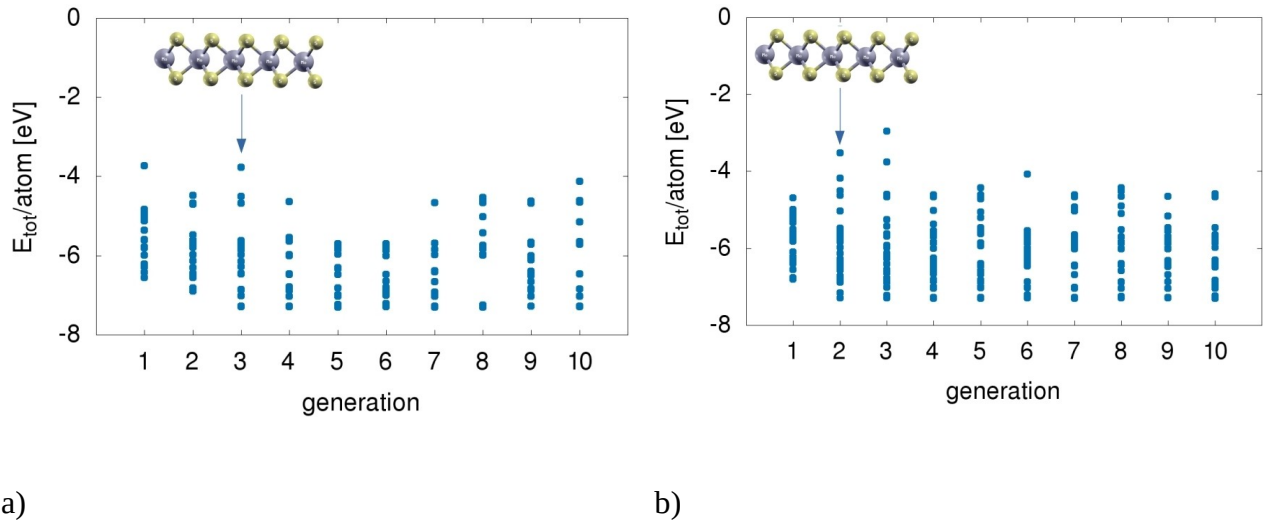

a)                                                b)

Fig. 2 Total energy per atom distribution for MoS$_2$ vs generation number for a) 24 and b) 48 individuals per generation.

III. CoGa$_2$S$_4$ – a layered material

As a next test, a more complicated stoichiometry has been chosen, namely CoGa$_2$S$_4$. The family of ternary dichalcogenides (TDs) with stoichiometry MGa$_2$X$_4$ (M-metal, X=S,Se,Te) has been recently reported [32] as very promising new 2D magnetic materials, with T$_C$ close to room temperature for CoGa$_2$X$_4$. These materials, as other TDs, have layered structure, where a 1T-like structure carrying the metal atom occupies the middle layer while two other layers are made of non-metallic ions (Fig. 3a). To check, whether such a structure can be obtained with our method (and is the most stable) we have conducted a series of calculations with two differences from the MoS$_2$ case. First, since the data for TDs are not present in the `c2db`, we have trained NN model with all the structures from this database. Second, we have added a new genetic operation which instead of single atoms, switches distinct layers, where a layer, lying on the XY plane, is defined as a set of at least two atoms having z coordinate close within some tolerance (here set to 0.1 Å).

As can be seen from Fig. 3 and Tab. 2, $CoGa_2S_4$ is highly polymorphic material. One of the main advantages of the evolutionary approach is the fact, that one obtains a set of structures, which can be sorted by the total energy. In this case, the most bound system is $CoGa_2S_4$ in 1T conformation (Fig. 3a), which is consistent with [32]. Also, three other formations presented in [32] have been obtained, namely (I)-T, (II)-T and (III)-T. Interestingly, fifth structure has been obtained (Fig. 3b), which is different from the other four – it is in fact a bilayer, which consist of $CoS_2$ and $Ga_2S_2$ layers, separated by distance of 2.9 Å. Also, in this case the Galium atoms form bounds. All four structures (except for 1T) are very close in energy (Tab. 2), while 1T lies in the clear energy minimum. The number of individuals in each generation has been set to 24, and the 1T structure appeared in 8$^{th}$ generation.

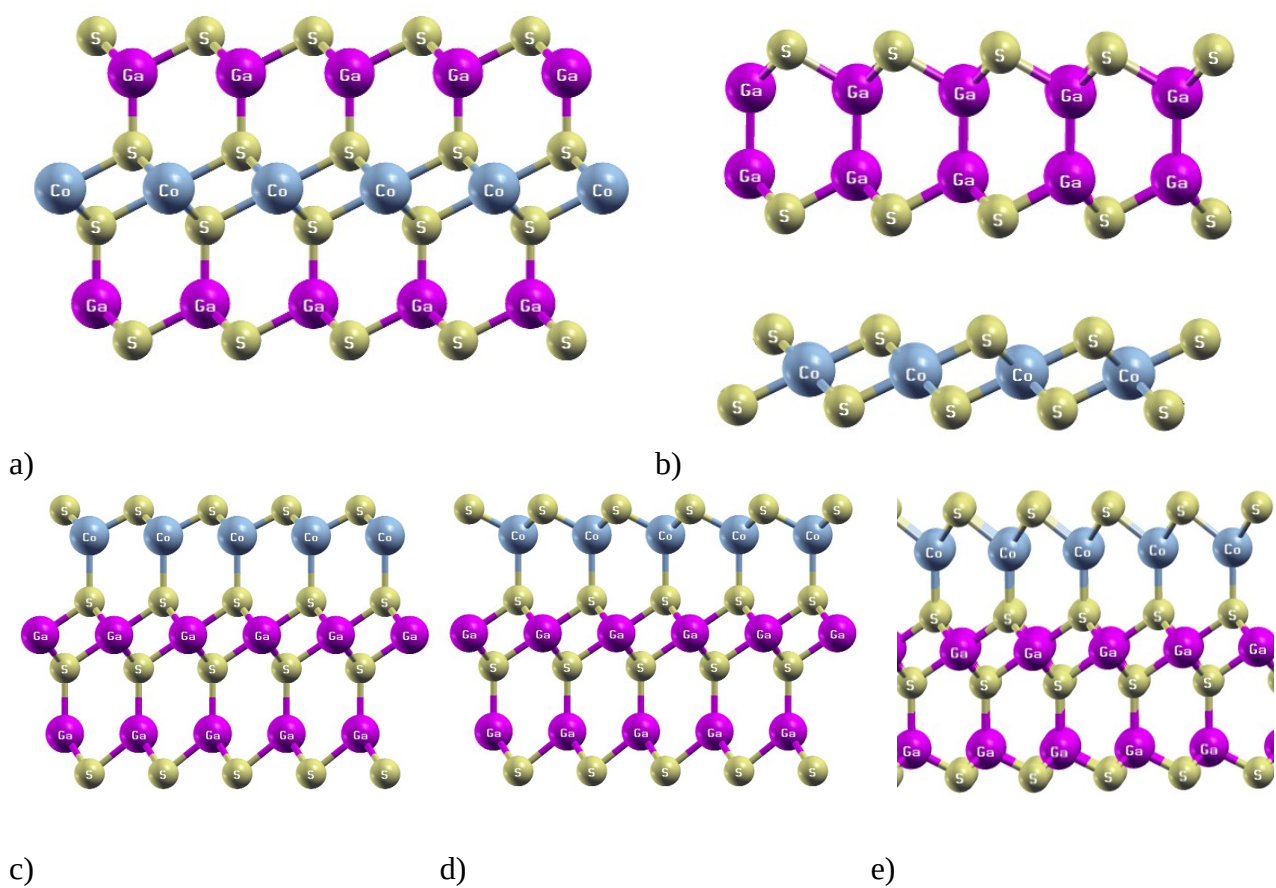

a) b) c) d) e)

Fig. 3 Conformations of $CoGa_2S_4$ structure: a) 1T, b) bilayer, c) (I)-T, d) (II)-T, e) (III)-T

|  | a [Å] | $E_{tot}$/atom [eV] |
|---|---|---|
| $CoGa_2S_4$ 1T | 3.62 | 4.77 |
| $CoGa_2S_4$ bilayer | 3.47 | 4.68 |
| $CoGa_2S_4$ (I)-T | 3.21 | 4.66 |
| $CoGa_2S_4$ (II)-T | 3.21 | 4.66 |
| $CoGa_2S_4$ (III)-T | 3.21 | 4.66 |

Table 2 Lattice constant and total energy values for different conformations of $CoGa_2S_4$

## IV. 2D boron structures on a substrate

The above cases focused on the prediction of atomic properties of pristine structures (i.e in a vacuum). In an experiment, the 2D structures are usually grown on a substrate. To address that type of problem, we have added a functionality to the code, which allows one to generate and evaluate structures on a given substrate, taking boron sheets on SiC as an example.

According to theoretical predictions, based on global optimization techniques [33] 2D monolayer of boron (known as borophene) is a polymorphic material, having many stable formations. Only recently, few-layer two-dimensional honeycomb boron has been studied by *ab-initio* methods [34], suggesting, that structures built of 3 or 4 layers are stable, with significant binding energy. Moreover, such structures are semiconductors with band gaps susceptible to strain, exerted for example by the presence of a substrate.

In [34], the pristine structures were optimized with the use of `USPEX`, and then the resulting few-layered boron sheets where put on SiC substrate. Here, we started with a substrate unit cell and prepared the initial population as a set of random boron layers on SiC. The stoichiometry of this system was ($B_6Si_6C_6$), where we have taken a unit cell of hexagonal SiC with a lattice constant of 3.09 Å as a basis. The rest of the algorithm have been realized as in the previous cases, although all the genetic operations have been conducted only on boron atoms and layers, leaving the SiC structure intact. As can be seen from Fig 4a, the resulting structure is identical to the one obtained in [34]. Namely, the boron atoms are arranged in layers with each layer shifted by a vector of (a/4,a/4) with respect to the lower one where a is a lattice constant of a SiC substrate. The final structure appeared in the structure pool in the 5$^{th}$ generation (Fig. 4b).

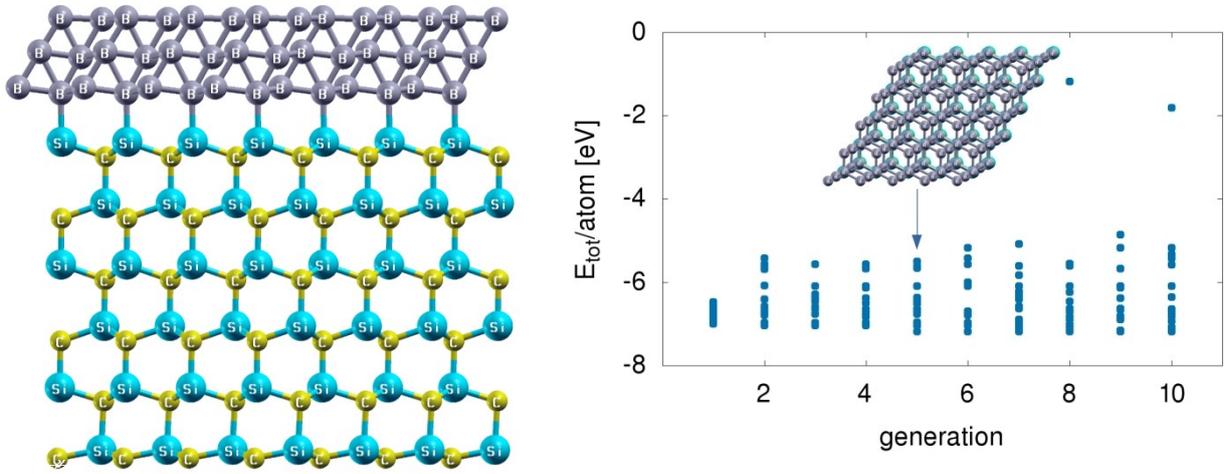

a)          b)

Fig. 4 Few-layer boron on a SiC substrate, a) structure side view, b) Total energy per atom distribution vs generation number and structure top view.

## V. $Cr_2Te_3$ – new stable 2D magnetic material

Two-dimensional ferromagnetic (FM) materials are at present one of the most promising candidates for nanoscale spintronic devices. Very recently, FM $Cr_2Ge_2Te_6$ bilayer [35], as well as $CrI_3$ [36] and $Fe_3GeTe_2$ [37] monolayers have been successfully synthesized. However, the Curie temperature ($T_C$) of these intrinsic 2D FM materials lies far below room temperature because of the weak ferromagnetic super-exchange interaction, preventing them from most applications. So, the pursuit of room temperature 2D FM materials continues. One of the most promising candidates is the family of 2D hexagonal $Cr_3X_4$ (X=S,Se,Te). According to recent theoretical predictions, $Cr_3Se_4$ and $Cr_3Te_4$ can have $T_C$ as high as, respectively, 370K and 460K [38].

Since bulk materials $Cr_2Te_3$, $Cr_3Te_4$, and $Cr_5Te_6$ are known for their magnetic properties for a very long time [39], we decided to look for 2D material with stoichiometry $Cr_2Te_3$, as it has never been addressed in the literature. The calculations have been conducted for 20 generations, 24 individuals in each. The most adapted structure turned out to be a hexagonal one (Fig. 5) with P1 symmetry group P1. Cr atoms form two layers with three layers of Te atoms, alternately. The first occurrence of the final structure has been detected in generation no. 7 (Fig. 6a). The resulting total energy is equal to -5.91 eV/atom. Also, the phonon spectrum of 2D $Cr_2Te_3$ has been calculated [40] with the frozen phonon method [41], as implemented in the `phonopy` code [42]. As can be seen from Fig 6b, phonon branches have no imaginary frequencies, suggesting the structure's dynamic stability.

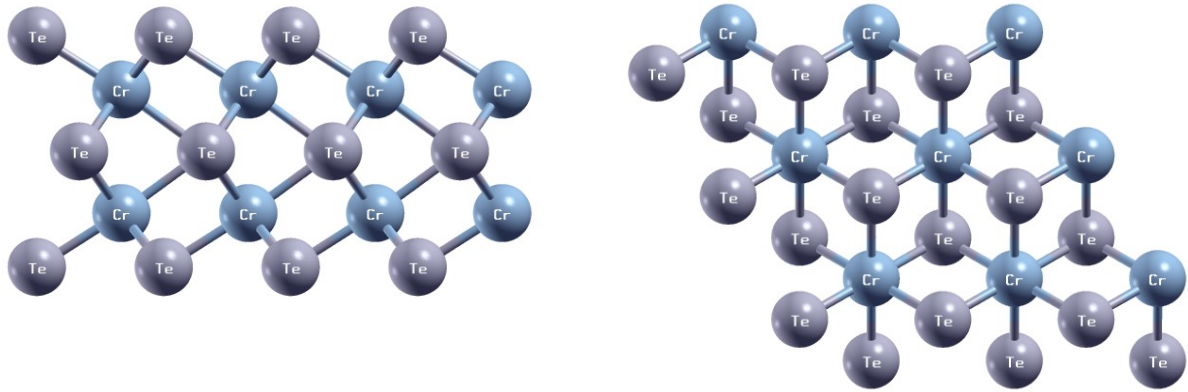

a) b)

Fig. 5 Atomic structure of 2D $Cr_2Te_3$: a) side view, b) top view

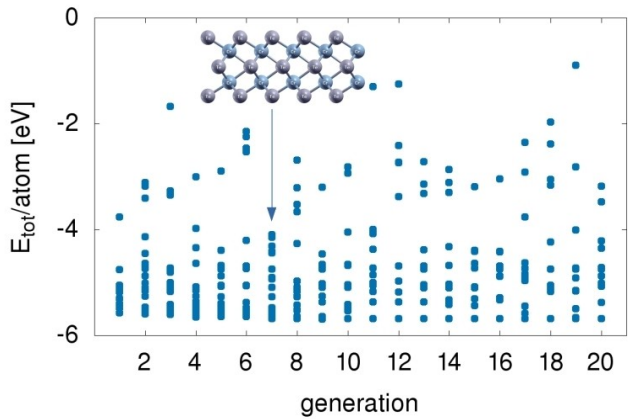 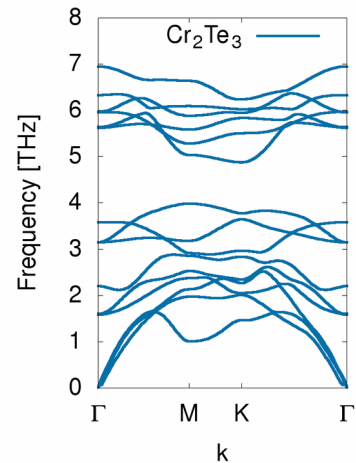

a) b)

Fig. 6 Results for 2D $Cr_2Te_3$: a) Total energy per atom distribution vs generation number, b) phonon spectrum

According to our calculations, the hexagonal $Cr_2Te_3$ will be a metallic material with FM alignment of spins and the total magnetic moment of 6.00µB/cell, concentrated almost solely on Cr atoms [43]. Our Monte Carlo (MC) simulation [44] based on 2D Heisenberg Hamiltonian [45] indicate, that the Curie temperature of $Cr_2Te_3$ will be close to 65K, which is much less than in the $Cr_3Te_4$. On the other hand, calculated magnetic anisotropy energy (MAE) equal to 1.55 meV/unit cell suggest the presence of an out-of-plane easy axis.

**Conclusions**

We have developed and implemented an algorithm of 2D structure prediction. The method, based on a combination of artificial neural networks and evolutionary approach has been tested for the cases of simple monolayer materials ($MoS_2$), layered materials ($CoGa_2S_4$) as well as 2D multilayers on a substrate (boron trilayer on SiC). In all the cases, the final structure was consistent with previous literature results. Finally, the code has been tested for the case of unknown structure, which resulted in new 2D stable magnetic material with the Curie temperature of 65K. The efficiency of the code can be considered as high since the lowest-energy structures appeared in the pool after at most eight generations.

**Acknowledgments**
Numerical calculations were supported in part by PL-Grid Infrastructure.

**References**
1. P. Hohenberg, W. Kohn, Phys. Rev. 136 (3B), B864 (1964).
2. S. Haastrup, M. Strange, M. Pandey, T. Deilmann, P. S. Schmidt, N. F. Hinsche, M. N. Gjerding, D. Torelli, P. M. Larsen, A. C. Riis-Jensen, J. Gath, K. W. Jacobsen, J. J. Mortensen, T. Olsen and K. S. Thygesen, 2D Materials 5, 042002 (2018).
3. I. Goodfellow, Y. Bengio, A. Courville "Deep Learning" MIT Press. ISBN 978-0-26203561-3 (2016).
4. H. K. D. H. Bhadeshia, ISIJ International. 39 (10): 966 (1999).
5. T. Bäck, D. Fogel, Z. Michalewicz, "Handbook of Evolutionary Computation", Oxford University Press, ISBN: 978-0750303927 (1997).
6. https://aflowlib.org
7. http://oqmd.org
8. https://cmr.fysik.dtu.dk/c2db/c2db.html
9. http://xtalopt.github.io/
10. A. R. Oganov, C. W. Glass J. Chem. Phys. 124, 24470 (2006); https://uspex-team.org/
11. J. Kennedy, R. C. Eberhart, "Swarm Intelligence" Morgan Kaufmann. ISBN 978-1-55860-595-4 (2001).
12. http://www.calypso.cn/
13. F. Oba and Y. Kumagai, App. Phys. Exp. 11, 060101 (2018).
13. W. Ye, Ch. Chen , Z. Wang, I.-H. Chu and S. Ping Ong, Nat. Comm. 9, 3800 (2018).
14. A. Majid, A. Khan, G. Javed and A. M. Mirza, Comp. Mat. Sci. 50, 363 (2010).
15. F. Musil, S. De, J. Yang, G. M. Day and M. Ceriotti, Chem. Sci. 9, 1289 (2017).
16. K. T. Schutt, F. Arbabzadah, S. Chmiela, K. R. Muller and A. Tkatchenko, Nat. Commun. 8, 13890 (2017).
17. F. A. Faber, L. Hutchison, B. Huang, J. Gilmer, S. S. Schoenholz, G. E. Dahl, O. Vinyals, S. Kearnes, P. F. Riley, and O. A. von Lilienfeld, J. Chem. Theory Comput. 13, 5255 (2017).
18. P. C. Jennings, S. Lysgaard, J. Strabo Hummelshoj, T. Vegge and T. Bligaard, npj Comput. Mater. 5, 46 (2019); E V. Podryabinkin, E. V. Tikhonov, A. V. Shapeev, and A. R. Oganov. Phys. Rev. B, 99, 064114 (2019).

19. C. Ashworth, Nat. Mat. Rev, 2, 18019 (2018).
20. B. Huang, G. Clark, E. Navarro-Moratalla, D. R. Klein, R. Cheng, K. L. Seyler, D. Zhong, E. Schmidgall, M. A. McGuire, D. H. Cobden, W. Yao, D. Xiao, P. Jarillo-Herrero and X. Xu, Nature 546, 270 (2017).
21. D. E. Goldberg, "Genetic Algorithms", Dorling Kindersley Pvt Ltd. ISBN: 978-8177588293 (2008).
22. G. Kresse and J. Hafner, Phys. Rev. B, 47:558, (1993); G. Kresse and J. Hafner, Phys. Rev. B, 49:14251, (1994); G. Kresse and J. Furthmüller, Comput. Mat. Sci., 6:15, (1996); G. Kresse and J. Furthmüller, Phys. Rev. B, 54:11169, (1996).
23. P. E. Blochl, Phys. Rev. B, 50:17953, (1994); G. Kresse and D. Joubert, Phys. Rev. B, 59:1758, (1999).
24. S. R. Bahn and K. W. Jacobsen, Comput. Sci. Eng., 4, 56, (2002);
https://wiki.fysik.dtu.dk/ase/index.html
25. S. Fredericks, D. Sayre and Q. Zhu, arXiv:1911.11123 (2019);
https://github.com/qzhu2017/PyXtal
26. P. W. Battaglia, J. B. Hamrick, V. Bapst, A. Sanchez-Gonzalez, V. Zambaldi, M. Malinowski, A. Tacchetti, D. Raposo, A. Santoro, Ryan Faulkner et. al., arXiv:1806.01261;
27. https://github.com/materialsvirtuallab/megnet
28. https://keras.io/
29. M. Brandbyge, J.-L. Mozos, P. Ordejon, J. Taylor, and K. Stokbro, Phys. Rev. B 65, 165401 (2002).
30. T. Ozaki, Phys. Rev. B 67, 155108 (2003).
31. VASP code was used with the PAW-PBE (GGA) pseudopotentials. All the structures were optimized until the forces inflicted on atoms were smaller than $10^{-5}$ eV/Å. The dense 30 x 30 x 1 k-points uniform grid was applied and the PW cutoff was set to 500 eV.
32. S. Zhang, R. Xu, W. Duan and Xiaolong Zou, Adv. Funct. Mater. 29, 1808380 (2019).
33. X.-F. Zhou, X. Dong, A.R. Oganov, Q. Zhu, Y. Tian, H.-T. Wang, Phys. Rev. Lett. 112, 085502 (2014).
34. K. Zberecki, Solid State Comm. 307, 113804 (2020).
35. C. Gong, L. Li, Z. Li, H. Ji, A. Stern, Y. Xia, T. Cao, W. Bao, C. Wang, Y. Wang, Z. Q. Qiu, R. J. Cava, S. G. Louie, J. Xia and X. Zhang, Nature, 546, 265 (2017).
36. B. Huang, G. Clark, E. Navarro-Moratalla, D. R. Klein, R. Cheng, K. L. Seyler, D. Zhong, E. Schmidgall, M. A. McGuire, D. H. Cobden, W. Yao, D. Xiao, P. Jarillo-Herrero and X. Xu, Nature 546, 270 (2017).
37. Z. Fei, B. Huang, P. Malinowski, W. Wang, T. Song, J. Sanchez, W. Yao, D. Xiao, X. Zhu, A. F. May, W. Wu, D. H. Cobden, J. H. Chu and X. Xu, Nat. Mater., 17, 778 (2018).
38. X. Zhang, B. Wang, Y. Guo, Y. Zhang, Y. Chen and J. Wang, Nanoscale Horiz., 4, 859 (2019).
39. A. F. Andersen, Acta Chem. Scand. 24, 3495 (1970).
40. For frozen phonon calculations the 4x4x1 supercell was used, after relaxation of a unit cell until the forces on each atom were smaller than $10^{-6}$ eV/Å.
41. K. Parlinski, Z. Q. Li, and Y. Kawazoe, Phys. Rev. Lett. 78, 4063 (1997).
42. A. Togo and I. Tanaka, Scr. Mater., 108, 1-5 (2015); https://phonopy.github.io/phonopy/
43. Detailed discussion of electronic and magnetic properties will be conducted in the separate paper.
44. R. F. L. Evans, W. J. Fan, P. Chureemart, T. A. Ostler, M. O. A. Ellis and R. W. Chantrell
J. Phys.: Condens. Matter 26, 103202 (2014); https://vampire.york.ac.uk/

45. D. Torelli, K. S. Thygesen and T. Olsen, 2D Mater. 6, 045018 (2019).
46. The code can be obtained freely from http://github.com/KrzysztofZb/NGOpt